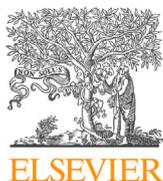

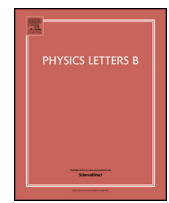

Letter

# Towards heavy-mass *ab initio* nuclear structure: Open-shell Ca, Ni and Sn isotopes from Bogoliubov coupled-cluster theory

A. Tichai [a,b,c], P. Demol [d], T. Duguet [e,d]

[a] *Technische Universität Darmstadt, Department of Physics, 64289 Darmstadt, Germany*
[b] *ExtreMe Matter Institute EMMI, GSI Helmholtzzentrum für Schwerionenforschung GmbH, 64291 Darmstadt, Germany*
[c] *Max-Planck-Institut für Kernphysik, Saupfercheckweg 1, 69117 Heidelberg, Germany*
[d] *KU Leuven, Instituut voor Kern- en Stralingsfysica, 3001 Leuven, Belgium*
[e] *IRFU, CEA, Université Paris-Saclay, 91191 Gif-sur-Yvette, France*

## ARTICLE INFO

Editor: B. Balantekin

## ABSTRACT

Recent developments in nuclear many-body theory enabled the description of open-shell medium-mass nuclei from first principles by exploiting the spontaneous breaking of symmetries within correlation expansion methods. Once combined with systematically improvable inter-nucleon interactions consistently derived from chiral effective field theory, modern *ab initio* nuclear structure calculations provide a powerful framework to deliver first-principle predictions accompanied with theoretical uncertainties. In this Letter, controlled *ab initio* Bogoliubov coupled cluster (BCC) calculations are performed for the first time, targeting the ground-state of all calcium, nickel and tin isotopes up to mass $A \approx 180$. While showing good agreement with available experimental data, the shell structure evolution in neutron-rich isotopes and the location of the neutron drip-lines are predicted. The BCC approach constitutes a key development towards reliable first-principles simulations of heavy-mass open-shell nuclei.

## 1. Introduction

The description of strongly correlated quantum systems poses a significant formal and computational challenge in various areas of many-body research. The exact solution of the many-body problem beyond the lightest nuclei becomes rapidly intractable due to the exponential scaling of quasi-exact diagonalization [1,2] and quantum Monte-Carlo [3–5] approaches. As a remedy, accessing nuclei with mass $A \geq 16$ is achieved by defining a simple, *e.g.*, mean-field, $A$-body reference state $|\Phi\rangle$ [6]. Truncating the expansion at a given order delivers a polynomially scaling method with system's size. In nuclear physics, the most commonly employed expansion schemes are many-body perturbation theory (MBPT) [7–13], self-consistent Green's function (SCGF) theory [14,15], the in-medium similarity renormalization group (IMSRG) [16–22] and coupled cluster (CC) theory [23–28].

Following such a strategy, heavy *closed-shell* nuclei (see Refs. [29–31]) have become recently accessible, all the way to $^{208}$Pb [32]. At the same time, nuclei away from shell closures still provide a formal and computational challenge. One way to overcome this limitation relies on valence-space (VS) techniques that build VS interactions from a variety of expansion methods for closed-shell nuclei [7,20,33,34] before proceeding to an exact diagonalization within that VS via modern shell-model codes [35,36]. In particular, the VS formulation based on the IMSRG has become the method of choice to describe medium-mass open-shell nuclei up to the lightest tin isotopes [29]. Still, VS approaches eventually suffer from the exponential scaling of the valence-space diagonalization, which makes it extremely difficult to push them to yet heavier open-shell nuclei. Thus, there have been extensive efforts to complement VS formulations with expansion methods building upon symmetry-broken reference states grasping strong static correlations associated with nuclear superfluidity and/or quadrupolar deformation [11,25,37,38]. In particular, breaking $U(1)$ symmetry associated to particle-number conservation leads to employing so-called Bogoliubov reference states. While such a paradigm was initially implemented within SCGF theory [37,39,40], it was later employed to design Bogoliubov MBPT (BMBPT) [11,12,41,42], Bogoliubov CC (BCC) [43] and Bogoliubov IMSRG [44] expansion methods. Alternatively, static correlations can be accounted for through multi-reference techniques that are built upon a non-product reference state [10,13,18,45–49]. In this context, the non-perturbative BCC theory is presently employed for the first time to address the heaviest open-shell nuclei ever computed *ab initio*.

---

* Corresponding author.
*E-mail addresses:* alexander.tichai@physik.tu-darmstadt.de (A. Tichai), pepijn.demol@kuleuven.be (P. Demol), thomas.duguet@cea.fr (T. Duguet).

## 2. Many-body framework

Particle-number-breaking many-body frameworks are formulated starting from a Bogoliubov reference state $|\Phi\rangle$ generated from the physical vacuum $|0\rangle$ through

$$|\Phi\rangle \equiv C \prod_k \beta_k |0\rangle,\tag{1}$$

where $C$ denotes a complex normalization constant. The state $|\Phi\rangle$ is a vacuum for the set of quasi-particle operators $\{\beta_k, \beta_k^\dagger\}$ generated from standard particle operators $\{c_l, c_l^\dagger\}$ via a unitary Bogoliubov transformation maintaining standard fermionic anti-commutation rules [50]. The variationally optimal Bogoliubov transformation is determined by solving Hartree-Fock-Bogoliubov (HFB) mean-field equations. In this work, rotational invariance is enforced to make $|\Phi\rangle$ an eigenstate of the total angular momentum $J^2$ and its projection $J_z$ with eigenvalues $J = 0$ and $M = 0$, respectively. Since particle-number symmetry is spontaneously broken by $|\Phi\rangle$, the Hamiltonian must be replaced by the (zero-temperature) grand potential $\Omega \equiv H - \lambda_N N - \lambda_Z Z$. The neutron (proton) chemical potential $\lambda_N$ ($\lambda_Z$) is tailored to constrain $|\Phi\rangle$ to carry the physical number of neutrons $N_0$ (protons $Z_0$) *on average*.

Building on such a reference state, Bogoliubov coupled-cluster theory [43] parametrizes the correlated ground-state wave function as $|\Psi\rangle \equiv e^{\mathcal{T}} |\Phi\rangle$ where the cluster operator $\mathcal{T} \equiv \mathcal{T}_1 + \mathcal{T}_2 + ...$ sums components inducing $2n$ quasi-particle excitations on top of $|\Phi\rangle$

$$\mathcal{T}_n \equiv \frac{1}{(2n)!} \sum_{k_1 ... k_{2n}} t_{k_1 ... k_{2n}} \beta_{k_1}^\dagger \cdots \beta_{k_{2n}}^\dagger,\tag{2}$$

where the set of tensors $t_{k_1 ... k_{2n}}$ constitute the unknown *cluster amplitudes*. The central quantity of the many-body formalism is the (non-Hermitian) similarity-transformed grand potential $\bar{\Omega} \equiv (e^{-\mathcal{T}} \Omega e^{\mathcal{T}})_c$, where the lower index 'c' stipulates the connected character of the operator [51]. Once cluster amplitudes have been determined by minimizing the BCC residual $\mathcal{R} = \langle \Phi | Q \bar{\Omega} | \Phi \rangle$, where $Q$ projects on the manifold of single, double, ... quasi-particle excitations of $\langle \Phi |$, the ground-state grand potential is computed as $\Omega_0 = \langle \Phi | \bar{\Omega} | \Phi \rangle$.

While BCC provides an exact parametrization of $|\Psi\rangle$, $\mathcal{T}$ must be truncated to make the problem numerically tractable. In this Letter, the BCC with singles and doubles (BCCSD) truncation is employed, i.e., $\mathcal{T}_{\mathrm{BCCSD}} \equiv \mathcal{T}_1 + \mathcal{T}_2$. The numerical implementation of BCCSD employs the quasi-linear form of the amplitude equation [43]. Rotational invariance is enforced on the amplitude equations thanks to angular-momentum-coupling techniques using the AMC program [52]. The iterative procedure to determine single and double amplitudes is initialized using first-order BMBPT wave functions and convergence is accelerated by employing the direct inversion of the iterative subspace (DIIS) approach [53]. The BCC iterations are carried out until all single and double residuals fall below $\|\mathcal{R}\| \leq 10^{-3}$ such that the ground-state energy and chemical potentials are stable at sub-keV level.

Since $|\Phi\rangle$ breaks U(1) symmetry, so does the *approximate* BCCSD wave function. Moreover, many-body correlations tend to induce a shift of the average particle number with respect to $|\Phi\rangle$ such that, e.g., $\langle \Psi_{\mathrm{BCCSD}} | N | \Psi_{\mathrm{BCCSD}} \rangle \neq N_0$. The BCCSD wave function is thus further constrained to carry the correct average neutron (proton) number,[1] i.e. $\langle \Phi | \tilde{N}(\tilde{Z}) | \Phi \rangle = N_0(Z_0)$, by readjusting the chemical potentials during the BCC iterations leading to the redefinition [54]

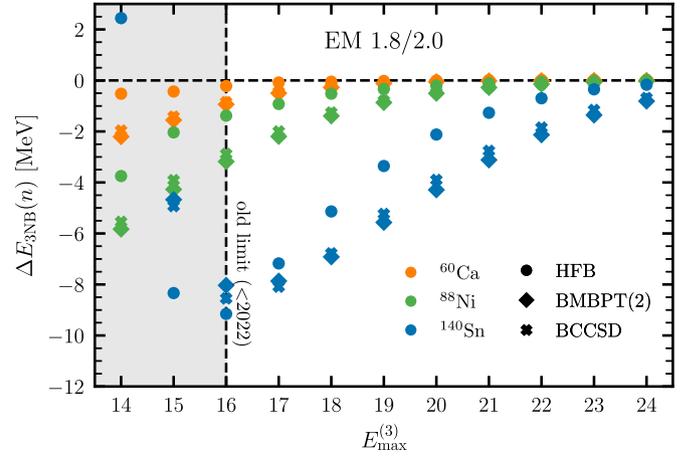

**Fig. 1.** Convergence of the ground-state energy as a function of $E_{\mathrm{max}}^{(3)}$ using $e_{\mathrm{max}} = 12$ with an oscillator frequency of $\hbar\Omega = 16$ MeV. In all cases the "EM1.8/2.0" interaction is employed.

$$\Omega_{\mathrm{shift}} \equiv \Omega + (\lambda_N + \lambda_N^{\mathrm{shift}})N + (\lambda_Z + \lambda_Z^{\mathrm{shift}})Z.\tag{3}$$

## 3. Interaction and model-space details

Numerical simulations of even-even Ca, Ni and Sn isotopes employ the "EM1.8/2.0" nuclear Hamiltonian [55] derived within chiral effective field theory [56–59]. While the three-nucleon (3N) interaction operator is fully accounted for at the HFB level, BCC equations are solved using the particle-number-conserving normal-ordered two-body (PNO2B) approximation [60]. The nuclear Hamiltonian is represented using a spherical harmonic oscillator (HO) single-particle basis truncated according to the principal quantum number $e_{\mathrm{max}} \equiv (2n + l)_{\mathrm{max}}$ of the last included state, where $n$ denotes the radial quantum number and $l$ the orbital angular momentum. To keep the number of 3N matrix elements tractable, an additional cut is performed by restricting three-body basis states to $e_1 + e_2 + e_3 \leq E_{\mathrm{max}}^{(3)}$. In all calculations the model space is truncated according to $e_{\mathrm{max}} = 12$ and $E_{\mathrm{max}}^{(3)} = 24$. A fully quantitative prediction of properties of neutron-rich systems eventually require a proper treatment of continuum effects (see e.g. Refs. [61,62]) that are not well resolved in the truncated harmonic oscillator single-particle basis employed in this work.

## 4. Many-body uncertainties

Gauging the quality of *ab initio* results necessitates an assessment of both Hamiltonian and many-body uncertainties. While the former, due to working at finite truncation order in the chiral power counting, is not addressed in this work, related discussions can be found in, e.g., Refs. [63–65]. The latter is schematically given by

$$\epsilon_{\mathrm{MB}} \equiv \epsilon_{\mathrm{FBS}} + \epsilon_{\mathrm{PNO2B}} + \epsilon_{\mathrm{3NB}} + \epsilon_{\mathrm{CC}} + \epsilon_{\mathrm{PNR}},\tag{4}$$

and is reported as a band in the figures below. The finite basis-size ($e_{\mathrm{max}}$) error $\epsilon_{\mathrm{FBS}}$ depends on the mass regime and varies among the studied isotopic chains: calcium (1%), nickel (1.5%) and tin (2%) at the employed $e_{\mathrm{max}} = 12$ value. Based on previous assessments in mid-mass closed-shell nuclei [66], the error $\epsilon_{\mathrm{PNO2B}}$ induced by omitting the residual normal-ordered 3N interaction is taken to be at the 2% level.

### 4.1. Three-body basis truncation error

The error $\epsilon_{\mathrm{3NB}}$ due to further truncating (via $E_{\mathrm{max}}^{(3)}$) the 3N basis has been the limiting factor to extend *ab initio* calculations to nuclei beyond Ni isotopes. Evaluating this error in details and eventually keeping it small enough up to neutron-rich tin isotopes is presently made possible by extending the novel three-body format of Ref. [31] to the context of $U(1)$-breaking expansion methods such as MBMPT and BCC.

---

[1] Since $N$ and $Z$ commute with $H$, the associated values in the BCC ground-state can, just as the energy, be evaluated using a projective approach.

The ground-state energy convergence is assessed in Fig. 1 by showing its evolution when incrementing $E_{\max}^{(3)}$ by one

$$\Delta E_{3NB}(n) \equiv E(E_{\max}^{(3)} = n) - E(E_{\max}^{(3)} = n - 1), \qquad (5)$$

where $E$ stands for the HFB, BMBPT(2) or BCCSD ground-state energy. For each isotopic chain a representative (neutron-rich) nucleus was chosen: $^{60}$Ca, $^{88}$Ni and $^{140}$Sn. The convergence depends on the nuclear mass. While $^{60}$Ca is reasonably well converged at $E_{\max}^{(3)} = 16$, there are still sizable contributions from discarded three-body matrix elements in heavier systems. In $^{140}$Sn, working with $E_{\max}^{(3)} = 16$ induces an error of the order of several tens of MeV, demonstrating the need for a significantly larger 3N basis when targeting nuclei with $A \gtrsim 80$ [31]. For sufficiently high values of $E_{\max}^{(3)}$, $\Delta E_{3NB}$ follows an empirically observed geometric progression $\Delta E_{3NB}(n) \sim q^n$. Hence the residual uncertainty can be estimated from the value obtained for the largest available $E_{\max}^{(3)} = 24$ times $q/(1-q)$, which for presently investigated systems amounts to a few tens of keV on total binding energies.

### 4.2. Many-body truncation error

The error $\epsilon_{CC}$ relates to the truncation of the BCC expansion, i.e. to the omission of high-rank cluster operators. At the BCCSD level, the error is essentially carried by missing triple excitations. In this respect, available VS-IMSRG(2) results [22] obtained in open-shell calcium isotopes with the same Hamiltonian constitute a reference since that static correlations are fully accounted for via the diagonalization within the valence shell. Consequently, a model is presently built for $\epsilon_{CC}$ by scaling the difference between BCCSD and VS-IMSRG(2) binding energies in (open-shell) Ca isotopes (see Fig. 2) with the BCCSD *correlation* energy. The former is systematically attractive and indeed shown to amount to about 8% of the latter along open-shell Ca isotopes. Eventually, $\epsilon_{CC}$ dominates the many-body error budget (Eq. (4)). With this model at hand, $\epsilon_{CC}$ can be propagated to Ni and Sn isotopes based on our sole BCCSD results. Of course, the goal in the future is to avoid the use of this rough error model by explicitly including (perturbative) triple corrections in BCC calculations.

A similar error model is also built for two-neutron separation energies ($S_{2n}$). Indeed, the difference between BCCSD and VS-IMSRG $S_{2n}$ is shown to be well approximated by taking 0.1% of the BCCSD correlation energy in open-shell Ca isotopes (see Fig. 2).

### 4.3. Particle-number violation error

The error $\epsilon_{PNR}$ is associated with the unphysical residual breaking of particle-number symmetry carried by the approximate BCCSD solution. This attractive contribution, presently estimated for each studied nucleus via effective field theory [67], is characterized in Sec. 6.

## 5. Numerical results

### 5.1. Calcium isotopic chain

Results of *ab initio* calculations along the calcium isotopic chain ($Z = 20$) are reported in Fig. 2. For the employed Hamiltonian 60-70% of the total binding energy is captured at the HFB level while the remaining is due to many-body correlations. Due to the softness of the "EM1.8/2.0" Hamiltonian, BMBPT(2) results are in good agreement with BCCSD. Eventually, BCCSD energies typically underestimate known experimental masses ($A \leq 58$) by about $5 - 13$ MeV ($1.7 - 2.9\%$). As testified by available VS-IMSRG(2) results [22], missing triple excitations would bring BCC calculations consistently closer to the data.

Two-neutron separation energies

$$S_{2n}(N, Z) \equiv E(N, Z) - E(N - 2, Z), \qquad (6)$$

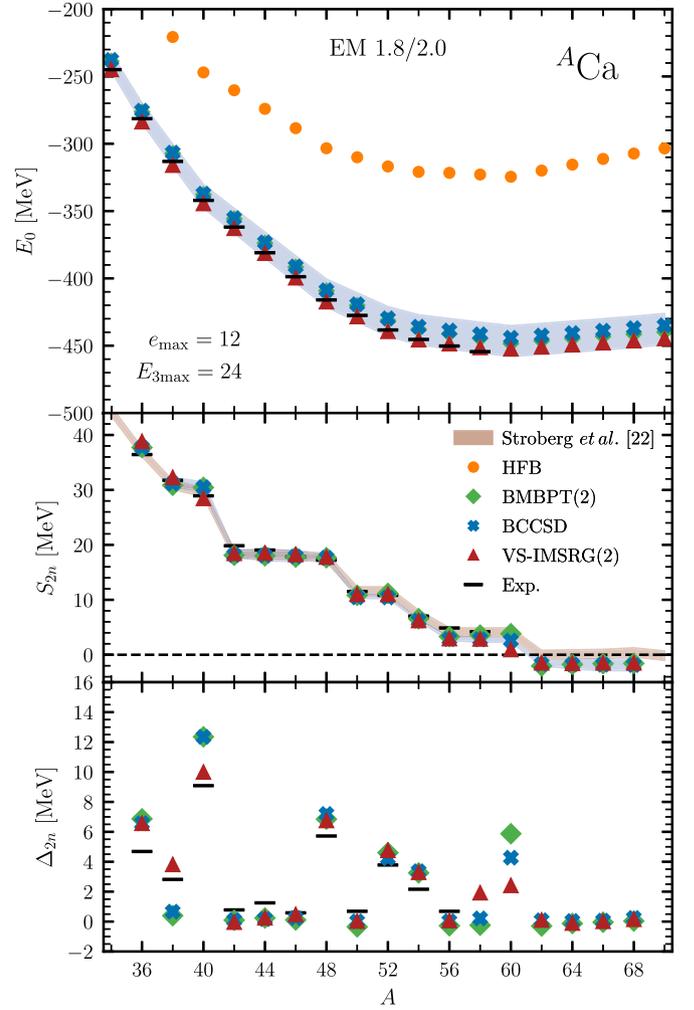

**Fig. 2.** Ground-state energy (top), two-neutron separation energy (middle) and two-neutron neutron shell gap (bottom) along the calcium isotopic chain using HFB (circle), BMBPT(2) (diamond), BCCSD (cross) and VS-IMSRG(2) (triangle). Calculations are carried out using $e_{\max} = 12$ and $E_{3\max}^{(3)} = 24$ with an oscillator frequency of $\hbar\Omega = 16$ MeV. In all cases the "EM1.8/2.0" Hamiltonian is employed. Experimental values are shown as black bars [70]. Bayesian uncertainties for $S_{2n}$ values from VS-IMSRG are taken from Ref. [22].

are also displayed in Fig. 2. Available experimental values [68,69] are reproduced within the 1–2 MeV many-body uncertainty. The drops at $^{52,54}$Ca indicate shell closures at neutron number $N = 32, 34$. Furthermore, the location of the neutron dripline ($S_{2n}(N, Z) < 0$) is predicted at $A = 60$ from both BMBPT(2) and BCCSD in correspondence with the closing of the $\nu 1f_{5/2}$ single-particle shell. These finding are consistent with VS-IMSRG(2) results.

Nuclear stability is commonly assessed from the excitation energy of the first $2_1^+$ state and its electromagnetic transition to the ground state. It can also be quantified from nuclear masses by computing the so-called two-neutron shell gap

$$\Delta_{2n}(N, Z) \equiv |S_{2n}(N, Z)| - |S_{2n}(N - 2, Z)|, \qquad (7)$$

a sudden increase of which indicates an extra stability associated in a mean-field picture with a closed-shell nucleus displaying a large Fermi gap. The bottom panel of Fig. 2 reveals pronounced kinks at $N = 20, 28$ and less pronounced kinks at $N = 32, 34$ in agreement with experimental data. Eventually, BMBPT(2) and BCCSD predict a significant shell closure at $N = 40$ that is however less pronounced in VS-IMSRG(2) calculations. This feature remains to be elucidated experimentally.

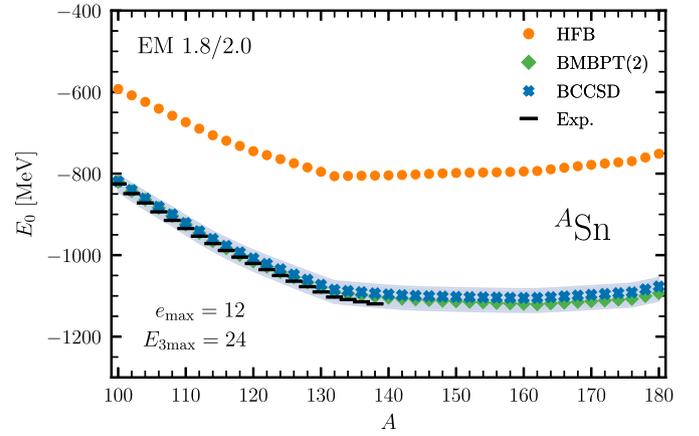

**Fig. 4.** Ground-state energy for $^{100-180}$Sn. Results from HFB, BMBPT(2) and BCCSD are compared to experimental data. Model-space parameters are identical to the ones employed in Figs. 2 and 3.

### 5.3. Tin isotopic chain

Finally, the most challenging tin ($Z = 50$) isotopic chain is targeted all the way to $A = 180$. While light tin isotopes were studied via VS-IMSRG calculations based on $E_{max}^{(3)} = 16$ [29], heavier isotopes are presently addressed via BMBPT and BCC calculations with $E_{max}^{(3)} = 24$. Based on the error model built in Ca isotopes, Fig. 4 demonstrates that BMBPT(2) and BCCSD binding energies are consistent with experiment up to $A = 138$ [70], although with a significant uncertainty due to missing correlations.

In Fig. 5, $S_{2n}$ from BCCSD calculations are displayed for three different values of the harmonic oscillator frequency around the empirical minimum of $\hbar\Omega = 12$ MeV. While $\epsilon_{FBS}$ was globally estimated to be around 2% along the tin isotopic chain, the use of three different $\hbar\Omega$ values testifies that such a basis truncation error is enhanced in the most neutron-rich isotopes and of the same order as $\epsilon_{CC}$ dominating the uncertainty band.[2] While BCCSD $S_{2n}$ are in good agreement with experiment up to $^{132}$Sn, they are significantly too low from $^{134}$Sn till $^{138}$Sn.[3] The difference to the data relates to an overestimation of the $N = 82$ magic shell gap and is inconsistent with our many-body error estimate. Consequently, it is probably attributable to the Hamiltonian uncertainty.

The flat evolution of BCCSD $S_{2n}$ beyond $^{140}$Sn leads, within estimated many-body uncertainties, to a large uncertainty on the location of the predicted neutron drip-line: $A \in [140 - 162]$. In spite of its large span, the predicted interval is inconsistent with the energy density functional (EDF) prediction, $A \in [172 - 176]$, from Ref. [74] (partially) accounting for systematic and statistical uncertainties of the EDF method. The location of the BCCSD prediction at significantly smaller neutron numbers clearly originates from the too large drop at $N = 82$. Consequently, the predicted interval should be enlarged to become consistent with the EDF prediction once the Hamiltonian uncertainty is included. Eventually, the uncertainty should be drastically reduced by including triple excitations and going to larger $e_{max}$ on the one hand and by reducing the (presently omitted) Hamiltonian uncertainty on the other hand.

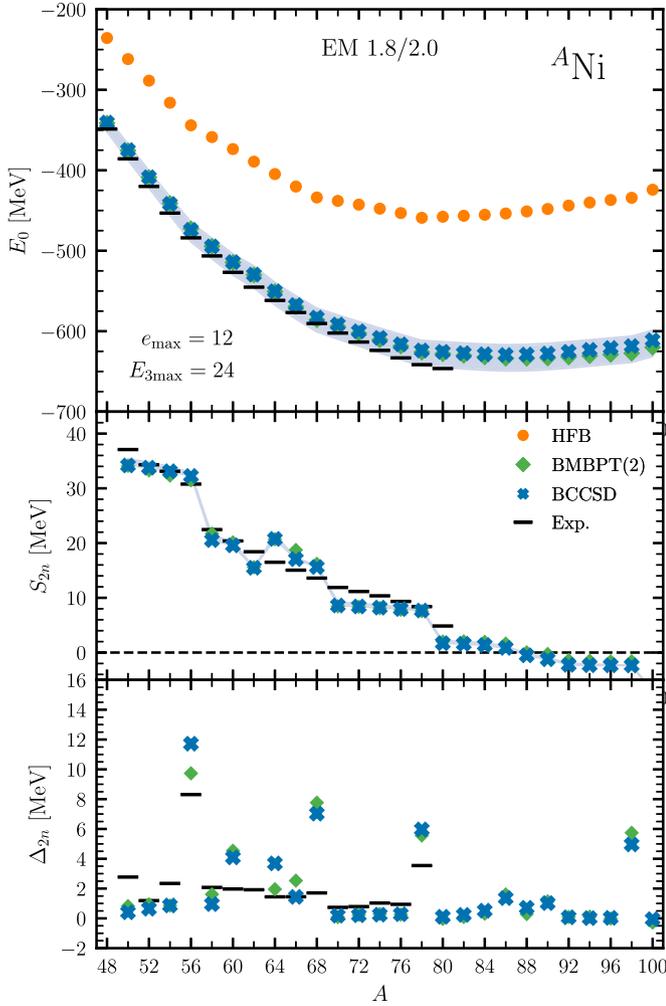

**Fig. 3.** Similar to Fig. 2 for nickel isotopes.

### 5.2. Nickel isotopic chain

Fig. 3 shows that predicted ground-state energies are consistent with experimental values up to the most neutron-rich observed isotope ($^{78}$Ni). The systematic underbinding is again due to currently discarded triple excitations. Neutron-rich nickel isotopes ($N = 50$ and beyond) provide a prime example for the limitations of valence-space techniques. In $^{78}$Ni the use of a $0\hbar\Omega$ space entails FCI dimensions of $\dim(\mathcal{H}_A) \sim 2 \cdot 10^{10}$ that cannot be handled by the most advanced shell-model code without severely truncating the configuration basis [71–73]. Hence the use of no-core approaches such as BCC provide a scalable solution in cases where the VS dimension becomes intractable and/or truncations of the shell model basis induce sizeable uncertainties.

The predicted dripline differs by two neutrons for BMBPT(2) ($^{88}$Ni) and BCCSD ($^{86}$Ni): the slightly positive BMBPT(2) value $S_{2n}(88, 28) = 35$ keV is contrasted by the BCC prediction $S_{2n}(88, 28) = -388$ keV yielding $^{86}$Ni as the last bound nickel isotope. While convergence with respect to model-space parameters is guaranteed, the relaxation of the many-body truncation and a more adequate treatment of the particle continuum may slightly impact the predicted dripline location. BMBPT(2) and BCCSD two-neutron shell gaps further predict magicity at $N = 28, 40, 50$. While this is confirmed experimentally for $N = 28$ and $N = 50$ [72], it is not the case at $N = 40$. A more definite statement requires the inclusion of $\mathcal{T}_3$ contributions and a systematic variation of the employed chiral Hamiltonian.

---

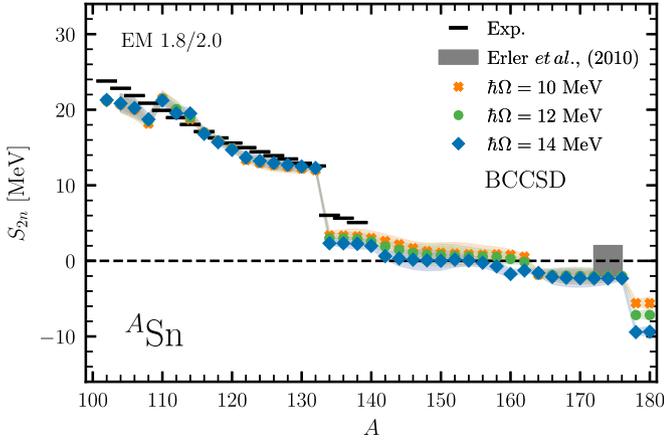

**Fig. 5.** BCCSD two-neutron separation energies for $^{100-180}$Sn using three different values for the harmonic oscillator frequency around the empirical minimum of $\hbar\Omega = 12$ MeV are compared to experimental data. The gray box locates the drip-line prediction from energy density functional calculations (including statistical uncertainty from model parameters) of Ref. [75].

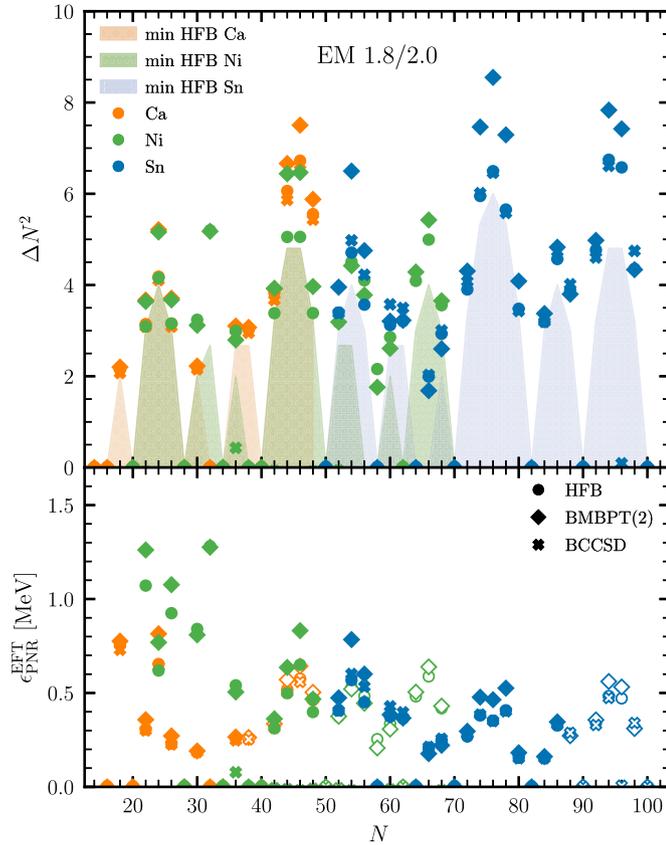

**Fig. 6.** Neutron-number variance (top) and leading-order EFT estimate of the energy gain from neutron-number restoration (bottom) for calcium, nickel and tin isotopic chains at HFB, BMBPT(2) and BCCSD truncation level. The low-energy scale $a^{-1}$ entering the EFT estimate [67] is taken from experimental data (open symbols are obtained using the last experimentally known open-shell isotope, i.e., $^{56}$Ca, $^{76}$Ni, $^{136}$Sn). Model-space parameters are identical to Figs. 2–4.

## 6. Particle-number breaking and restoration

While the HFB reference state breaks particle-number conservation in open-shell systems, Fig. 6 displaying HFB, BMBPT(2) and BCCSD neutron-number variances $\Delta N^2 \equiv \langle (N - \langle N \rangle)^2 \rangle$ demonstrates that the

"EM1.8/2.0" Hamiltonian induces only weak pairing correlations at the mean-field level. Indeed the HFB variance is very close to the minimal boundary corresponding to the zero-pairing limit [76] along Ca, Ni and Sn isotopic chains except, typically, for the most neutron-rich open-shell isotopes. Because $|\Psi\rangle$ is an eigenstate of $N$, one expects that the symmetry violation is reduced by improving on the BMBPT/BCC truncation. While the neutron-number variance essentially increases in BMBPT(2) calculations, it typically only displays a slight reduction at the BCCSD level compared to HFB. This shows both that the variance reduction is not monotonic in BMBPT [42,77] and that the symmetry restoration cannot effectively be achieved via the summation of low-rank individual excitations. This calls for an explicit particle-number-projection in open-shell nuclei. While this is a standard procedure at the mean-field level [50], extensions to correlation expansions have been explored only recently [28,78–81]. In absence of an explicit projection, an effective field theory (EFT) is presently used to estimate, at each truncation order, the missing energy contribution $\epsilon_{PNR} \sim -a^{-1}\Delta N^2$, where $a$ is a characteristic low-energy scale [67]. The bottom panel of Fig. 6 illustrates that this (non-size-extensive) contribution decreases from about 1 MeV in the Ca-Ni region to about 0.5 MeV in Sn isotopes with significant local variation associated with the filling of successive neutron shells.

## 7. Conclusions and perspectives

In this Letter, large-scale ab initio non-perturbative Bogoliubov coupled cluster (BCC) calculations were performed for the first time to systematically describe ground-state energies along Ca, Ni and Sn isotopic chains from first principles. These constitute the heaviest open-shell nuclei ($A \sim 180$) ever computed within an ab initio framework. Such an achievement was made possible by extending to $U(1)$-breaking expansion methods such as BCC and the optimized storage scheme of chiral three-body forces recently employed in the computation of doubly closed-shell $^{132}$Sn [31] and $^{208}$Pb [32].

Various research directions open up for the near future. First the BCC truncation will be relaxed to incorporate the effect of triple excitations [23,25,82] and reduce the dominating many-body uncertainty. To further reduce uncertainties in heavy neutron-rich isotopes, calculations will be pushed to $e_{max} > 12$. To cope with the increased dimensionality, this will be combined with basis-optimization techniques accelerating the convergence of many-body observables [25,83,84]. Moreover, the formulation of an equation-of-motion [51] BCC extension will provide access to the spectroscopy of even-even and even-odd open-shell nuclei beyond $A = 100$. Eventually, the implementation of a particle-number restoration [79,80] is envisioned to explicitly account for $\epsilon_{PNR}$ as well as to compute rotational excitations and pair-transfer probabilities.

### Declaration of competing interest

The authors declare that they have no known competing financial interests or personal relationships that could have appeared to influence the work reported in this paper.

### Data availability

Data will be made available on request.

### Acknowledgements

We thank P. Arthuis, G. Hagen, T. Miyagi and A. Schwenk fur useful discussions. We further thank G. Hagen for helping us benchmark our BCC code, and T. Miyagi for providing us with VS-IMSRG results as well for generating chiral input matrix elements. A.T. greatfully acknowledges the help of M. Frosini in benchmarking our HFB solver. This work was supported in part by the Deutsche Forschungsgemeinschaft (DFG, German Research Foundation) – Projektnummer 279384907 – SFB 1245, by the European Research Council (ERC) under the European Union's Horizon 2020 research and innovation programme (Grant Agreement No. 101020842), and by Research Foundation Flanders (FWO, Belgium, grant 11G5123N). Computations were in part performed with an allocation of computing resources at the Jülich Supercomputing Center.